\documentstyle[12pt,psfig,epsfig,citesort,enumerate]{siam}
\newcommand{\EXP}{2^{n^{O(1)}}}
\newcommand{\Xomit}[1]{}
\newcommand{\code}[1]{\mbox{\rm code}_\pi({#1},\lambda)}
\newcommand{\algo}[1]{\mbox{\rm encode}_\pi({#1})}
\renewcommand{\index}[1]{\mbox{index}_\pi({#1})}
\newcommand{\gin}{G_{\mbox{\scriptsize in}}}
\newcommand{\gout}{G_{\mbox{\scriptsize out}}}
\newcommand{\dfs}{\mbox{dfs}}
\newcommand{\setof}[1]{\{{#1}\}}
\newtheorem{fact}{Fact}

\newcommand{\ceil}[1]{\lceil{#1}\rceil}

\begin{document}

\title{A Fast General Methodology for Information-Theoretically Optimal
Encodings of Graphs\thanks{A preliminary version appeared in {\em
Proceedings of the 7th Annual European Symposium on Algorithms}, 
Lecture Notes in Computer Science 1643 (1999), pp.~540--549.}}
\author{ 
Xin He\thanks{Department of Computer Science and Engineering, State
University of New York at Buffalo, Buffalo, NY 14260, USA.  Email:
xinhe@cse.buffalo.edu.}
\and
Ming-Yang Kao\thanks{Department of Computer Science, Yale University,
New Haven, CT 06250, USA.  Email: kao-ming-yang@cs.yale.edu. Research
supported in part by NSF Grant CCR-9531028.}
\and
Hsueh-I Lu\thanks{Institute of Information Science, Academia Sinica,
Taipei 115, Taiwan, ROC. Email: hil@iis.sinica.edu.tw. 
Part of the work is performed at
Department of Computer Science and Information
Engineering, National Chung-Cheng University, Chia-Yi 621, Taiwan,
ROC. Research supported in Part by NSC Grant NSC-89-2213-E-001-034.
}}

\date{\today}

\maketitle

\begin{abstract}
We propose a fast methodology for encoding graphs with
information-theoretically minimum numbers of bits.  Specifically, a
graph with property $\pi$ is called a {\em $\pi$-graph}.  If $\pi$
satisfies certain properties, then an $n$-node $m$-edge $\pi$-graph
$G$ can be encoded by a binary string $X$ such that (1) $G$ and $X$
can be obtained from each other in $O(n\log n)$ time, and (2) $X$ has
at most $\beta(n)+o(\beta(n))$ bits for any continuous super-additive
function $\beta(n)$ so that there are at most
$2^{\beta(n)+o(\beta(n))}$ distinct $n$-node $\pi$-graphs.  The
methodology is applicable to general classes of graphs; this paper
focuses on planar graphs. Examples of such $\pi$ include all
conjunctions over the following groups of properties: 
(1) $G$ is a planar graph or a plane graph; 
(2) $G$ is directed or undirected; 
(3) $G$ is triangulated, triconnected, biconnected, merely connected, 
or not required to be connected; 
(4) the nodes of $G$ are labeled with labels 
from $\setof{1,\ldots, \ell_1}$ for $\ell_1\leq n$;
(5) the edges of $G$ are labeled with labels 
from $\setof{1,\ldots, \ell_2}$ for $\ell_2\leq m$; and
(6) each node (respectively, edge) of $G$ has at most $\ell_3=O(1)$ 
self-loops (respectively, $\ell_4=O(1)$ multiple edges).
Moreover, $\ell_3$ and
$\ell_4$ are not required to be $O(1)$ for the cases of $\pi$ being a
plane triangulation.  These examples are novel applications of small
cycle separators of planar graphs and are the only nontrivial classes
of graphs, other than rooted trees, with known polynomial-time
information-theoretically optimal coding schemes.
\end{abstract}

\Xomit{


\begin{AMSMOS}
05C10, 05C30, 05C78, 05C85, 68R10, 65Y25, 94A15
\end{AMSMOS}

}

\section{Introduction}
Let $G$ be a graph with $n$ nodes and $m$ edges.  This paper studies
the problem of {\em encoding} $G$ into a binary string $X$ with the
requirement that $X$ can be {\em decoded} to reconstruct $G$. We
propose a fast methodology for designing a coding scheme such that the
bit count of $X$ is information-theoretically optimal.  Specifically,
a function $\beta(n)$ is {\em super-additive} if
$\beta(n_1)+\beta(n_2)\leq\beta(n_1+n_2)$. A function $\beta(n)$ is {\em
continuous} if $\beta(n+o(n))=\beta(n)+o(\beta(n))$. For example,
$\beta(n)=n^c\log^dn$ is continuous and super-additive, for any
constants $c\geq 1$ and $d\geq 0$. The continuity and super-additivity
are closed under additions.  A graph with property $\pi$ is called a
{\em $\pi$-graph}. If $\pi$ satisfies certain properties, then we can
obtain an $X$ such that (1) $G$ and $X$ can be computed from each
other in $O(n\log n)$ time and (2) $X$ has at most
$\beta(n)+o(\beta(n))$ bits for any continuous super-additive function
$\beta(n)$ so that there are at most $2^{\beta(n)+o(\beta(n))}$
distinct $n$-node $m$-edge $\pi$-graphs. The methodology is applicable
to general classes of graphs; this paper focuses on planar graphs.

A {\em conjunction} over $k$ groups of properties is a boolean property
$\pi_1\wedge\cdots\wedge\pi_k$, where $\pi_i$ is a property in the
$i$-th group for each $i=1,\ldots,k$. 
Examples of suitable $\pi$ for our methodology 
include every conjunction over the 
following groups: 
\begin{enumerate}[\qquad F1.]
\item\label{f1}
$G$ is a planar graph or a plane graph.
\item\label{f2}
$G$ is directed or undirected.
\item\label{f3} 
$G$ is triangulated, triconnected, biconnected, merely connected, or not required to
be connected.
\item\label{f4}
The nodes of $G$ are labeled with labels from $\setof{1,\ldots, \ell_1}$ for
$\ell_1\leq n$.
\item\label{f5} 
The edges of $G$ are labeled with labels from $\setof{1,\ldots, \ell_2}$ for
$\ell_2\leq m$.
\item\label{f6} 
Each node of $G$ has at most $\ell_3=O(1)$ self-loops.
\item\label{f7}   
Each edge of $G$ has at most $\ell_4=O(1)$ multiple edges.
\end{enumerate}
Moreover, $\ell_3$ and $\ell_4$ are not required to be $O(1)$ for the
cases of $\pi$ being a plane triangulation.  For instance, $\pi$ can
be the property of being a directed unlabeled biconnected simple plane
graph.  These examples are novel applications of small cycle
separators of planar graphs
\cite{Miller86,LiptonT79}.  
Note that the rooted trees are the only other nontrivial class of
graphs with a known polynomial-time information-theoretically optimal
coding scheme, which encodes a tree as nested parentheses using
$2(n-1)$ bits in $O(n)$ time.

Previously, Tutte proved that there are $2^{\beta(m)+o(\beta(m))}$
distinct $m$-edge plane triangulations where $\beta(m)=
(\frac{8}{3}-\log_2 3)m + o(m)\approx1.08m + o(m)$ \cite{Tutte62} and
that there are $2^{2m+o(n)}$ distinct $m$-edge $n$-node triconnected
plane graphs that may be non-simple \cite{Tutte63b}.
Tur\'{a}n~\cite{turan84} used $4m$ bits to encode a plane graph $G$
that may have self-loops. Keeler and Westbrook~\cite{KW:encodings}
improved this bit count to $3.58m$.  They also gave coding schemes for
several families of plane graphs. In particular, they used $1.53m$
bits for a triangulated simple $G$, and $3m$ bits for a connected $G$
free of self-loops and degree-one nodes.  For a simple triangulated
$G$, He, Kao, and Lu~\cite{HeKL-sidm} improved the bit count to
$\frac{4}{3}m+O(1)$.  For a simple $G$ that is triconnected and thus
free of degree-one nodes, they \cite{HeKL-sidm} improved the bit count
to at most $2.835m$ bits.  This bit count was later reduced to at most
$\frac{3\log_23}{2}m+O(1)\approx 2.378m+O(1)$ by Chuang, Garg, He,
Kao, and Lu~\cite{ChuangGHKL98}.  These coding schemes all take linear
time for encoding and decoding, but their bit counts are not
information-theoretically optimal. For labeled planar graphs, Itai and
Rodeh \cite{IR82} gave an encoding of $\frac{3}{2} n \log n + O(n)$
bits.  For unlabeled general graphs, Naor \cite{naor90} gave an
encoding of ${\frac{1}{2}}n^2-n\log{n}+ O(n)$ bits.

For applications that require query support,
Jacobson~\cite{Jacobson89} gave a $\Theta(n)$-bit encoding for a
connected and simple planar graph $G$ that supports traversal in
$\Theta(\log{n})$ time per node visited. Munro and Raman~\cite{MR97}
improved this result and gave schemes to encode binary trees, rooted
ordered trees, and planar graphs. For a general planar $G$, they used
$2m+8n+o(m+n)$ bits while supporting adjacency and degree queries in
$O(1)$ time. Chuang {\em et al.}~\cite{ChuangGHKL98} reduced this bit
count to $2m+(5+\frac{1}{k})n+o(m+n)$ for any constant $k>0$ with the
same query support. The bit count can be further reduced if only
$O(1)$-time adjacency queries are supported, or if $G$ is simple,
triconnected or triangulated~\cite{ChuangGHKL98}.  For certain graph
families, Kannan, Naor and Rudich~\cite{KNR92} gave schemes that
encode each node with $O(\log n)$ bits and support $O(\log n)$-time
testing of adjacency between two nodes. For dense graphs and
complement graphs, Kao, Occhiogrosso, and Teng~\cite{kaot93.joa}
devised two compressed representations from adjacency lists to speed
up basic graph search techniques. Galperin and Wigderson \cite{GW83}
and Papadimitriou and Yannakakis~\cite{PH86.encode} investigated
complexity issues arising from encoding a graph by a small circuit
that computes its adjacency matrix.

Section~\ref{section:methodology} discusses the general encoding
methodology. Sections~\ref{section:triangulation}
and~\ref{section:planar} use the methodology to obtain
information-theoretically optimal encodings for various classes of
planar graphs. Section~\ref{sec_open} concludes the paper with some
future research directions.

\section{The encoding methodology}\label{section:methodology}
Let $|X|$ be the number of bits in a binary string $X$. Let $|G|$ be the
number of nodes in a graph $G$. Let $|S|$ be the number of elements,
counting multiplicity, in a multiset $S$.
\begin{fact}[see \cite{BCW90,Elias75}]\label{fact:aux}
Let $X_1,X_2,\ldots, X_k$ be $O(1)$ binary strings. Let
$n=|X_1|+|X_2|+\cdots+|X_k|$.  Then there exists an $O(\log n)$-bit
string $\chi$, obtainable in $O(n)$ time, such that given the
concatenation of $\chi,X_1,X_2,\ldots,X_k$, the index of the first
symbol of each $X_i$ in the concatenation can be computed in $O(1)$
time.
\end{fact}

Let $X_1+X_2+\cdots+X_k$ denote the concatenation of
$\chi,X_1,X_2,\ldots,X_k$ as in Fact~\ref{fact:aux}. We call $\chi$ the
{\em auxiliary binary string} for $X_1+X_2+\cdots+X_k$. 

A graph with property $\pi$ is called a {\em $\pi$-graph}.  Whether
two $\pi$-graphs are {\em distinct} or {\em indistinct} depends on
$\pi$. For example, let $G_1$ and $G_2$ be two topologically
non-isomorphic plane embeddings of the same planar graph.  If $\pi$ is
the property of being a planar graph, then $G_1$ and $G_2$ are two
indistinct $\pi$-graphs. If $\pi$ is the property of being a planar
embedding, then $G_1$ and $G_2$ are two distinct $\pi$-graphs.  Let
$\alpha$ be the number of distinct $n$-node $\pi$-graphs.  Clearly it
takes $\ceil{\log_2\alpha}$ bits to differentiate all $n$-node
$\pi$-graphs.  Let $\index{G}$ be an $\ceil{\log_2\alpha}$-bit
indexing scheme of the $\alpha$ distinct $\pi$-graphs.

Let $G_0$ be an input $n_0$-node $\pi$-graph.  Let
$\lambda=\log\log\log(n_0)$.  The encoding algorithm $\algo{G_0}$ is
merely a function call $\code{G_0}$, where the recursive function
$\code{G}$ is defined as follows:
\begin{tabbing}
\qquad\=function $\code{G}$\\
           \>\{\\
\>\quad\=if $|G|=O(1)$ or $|G|\leq \lambda$ then\\
\>\>\quad\=return $\index{G}$\\
\>        \>else\\
\>        \>\{\\
\>        \>\>compute $\pi$-graphs $G_1,G_2$, and a string $X$, from
              which $G$ can be recovered;\\
\>        \>\>return $\code{G_1}+\code{G_2}+X$;\\
\>        \>\}\\
\>\}
\end{tabbing}
Clearly, the code returned by algorithm $\algo{G_0}$ can be decoded to
recover $G_0$. For notational brevity, if it is clear from the context, the code
returned by algorithm $\algo{G_0}$ (respectively, function $\code{G}$)
is also denoted $\algo{G_0}$ (respectively, $\code{G}$).

Function $\code{G}$ {\em satisfies the separation property} if there
exist two constants $c$ and $r$, where $0\leq c <1$ and $r>1$, such
that the following conditions hold:
\begin{enumerate}[\qquad P1.]
\item\label{s1} $\max(|G_1|,|G_2|)\leq |G|/r$.
\item\label{s2} $|G_1|+|G_2|=|G|+O(|G|^c)$.
\item\label{s3} $|X|=O(|G|^c)$.
\end{enumerate}

Let $f(|G|)$ be the time required to obtain $\index{G}$ and $G$ from
each other. Let $g(|G|)$ be the time required to obtain $G_1,G_2,X$ from
$G$, and vice versa.
\begin{theorem}\label{theorem:main}
Assume that function $\code{G}$ satisfies the separation property; and
that there are at most $2^{\beta(n)+o(\beta(n))}$ distinct $n$-node
$\pi$-graphs for some continuous super-additive function $\beta(n)$.
\begin{enumerate}
\item\label{main1} $|\algo{G_0}|\leq\beta(n_0)+o(\beta(n_0))$ for any
$n_0$-node $\pi$-graph $G_0$.
\item\label{main2} If $f(n)=\EXP$ and $g(n)=O(n)$, then 
$G_0$ and $\algo{G_0}$ can be obtained from each other in $O(n_0\log
n_0)$ time.
\end{enumerate}
\end{theorem}
\begin{proof}
The theorem holds trivially if $n_0=O(1)$. For the rest of the proof
we assume $n_0=\omega(1)$, and thus $\lambda=\omega(1)$.
Many graphs may appear during the execution of $\algo{G_0}$. These
graphs can be organized as nodes of a binary tree $T$ rooted at
$G_0$, where (i) if $G_1$ and $G_2$ are obtained from $G$ by calling
$\code{G}$, then $G_1$ and $G_2$ are the children of $G$ in $T$, and
(ii) if $|G|\leq\lambda$, then $G$ has no children in $T$.
Further consider the multiset $S$ consisting of all graphs $G$ that are nodes
of $T$.  We partition $S$ into $\ell+1$ multisets
$S(0),S(1),S(2),\ldots,S(\ell)$ as follows. $S(0)$ consists of the
graphs $G$ with $|G|\leq\lambda$.  For $i\geq 1$, $S(i)$ consists of
the graphs $G$ with $r^{i-1}\lambda< |G|\leq r^i\lambda$. Let $G_0\in
S(\ell)$, and thus set $\ell=O(\log\frac{n_0}{\lambda})$.

Define $p=\sum_{H\in S(0)}|H|$. We first show
\begin{equation}\label{eq1}
|S(i)|< \frac{p}{r^{i-1}\lambda},
\end{equation}
for every $i=1,\ldots,\ell$. Let $G$ be a graph in $S(i)$.  Let
$S(0,G)$ be the set consisting of the leaf descendants of $G$ in $T$;
for example, $S(0,G_0)=S(0)$. By Condition~P\ref{s2},
$|G|\leq\sum_{H\in S(0,G)} |H|$.  By Condition~P\ref{s1}, no two graphs
in $S(i)$ are related in $T$.  Therefore $S(i)$ contains at most one
ancestor of $H$ in $T$ for every graph $H$ in $S(0)$. It follows that
$\sum_{G\in S(i)}|G|\leq\sum_{G\in S(i)}\sum_{H\in S(0,G)} |H|\leq p$.
Since $|G|> r^{i-1}\lambda$ for every $G$ in $S(i)$,
Inequality~(\ref{eq1}) holds.

\paragraph{\rm Statement 1}
Suppose that the children of $G$ in $T$ are $G_1$ and $G_2$. Let
$b(G)=|X|+|\chi|$, where $\chi$ is the auxiliary binary string for
$\code{G_1}+\code{G_2}+X$.  Let $q=\sum_{i\geq 1}\sum_{G\in
S(i)}b(G)$. Then, $|\algo{G_0}|=q+\sum_{H\in S(0)}|\code{H}|\leq
q+\sum_{H\in S(0)}(\beta(|H|)+o(\beta(|H|)))$. By the super-additivity
of $\beta(n)$, $|\algo{G_0}|\leq q+\beta(p)+o(\beta(p))$.  Since
$\beta(n)$ is continuous, Statement~1 can be proved by showing
$p=n_0+o(n_0)$ and $q=o(n_0)$ below.

By Condition~P\ref{s3}, $|X|=O(|G|^c)$. By Fact~\ref{fact:aux},
$|\chi|=O(\log |G|)$. Thus, $b(G)=O(|G|^c)$, and
\begin{equation}\label{eq2}
  q=\sum_{i\geq 1}\sum_{G\in S(i)}O(|G|^c).
\end{equation}
Now we regard the execution of $\algo{G_0}$ as a process of
growing $T$. Let $a(T)=\sum_{\mbox{\scriptsize$H$ is a leaf of
$T$}}|H|$. At the beginning of the function call $\algo{G_0}$, $T$
has exactly one node $G_0$, and thus $a(T)=n_0$. At the end of the
function call, $T$ is fully expanded, and thus $a(T)=p$. By 
Condition~P\ref{s2}, during the execution of
$\algo{G_0}$, every function call $\code{G}$ with $|G|>\lambda$
increases $a(T)$ by $O(|G|^c)$.  Hence
\begin{equation}\label{eq3}
  p=n_0+\sum_{i\geq 1}\sum_{G\in S(i)}O(|G|^c).
\end{equation}
Note that
\begin{equation}\label{eq4}
  \sum_{i\geq 1}\sum_{G\in S(i)}|G|^c
  \leq\sum_{i\geq 1}(r^i\lambda)^c{p}/{(r^{i-1}\lambda)}
  =p\lambda^{c-1}r\sum_{i\geq 1}r^{(c-1)i}
  =p\lambda^{c-1}O(1)
  =o(p).
\end{equation}
By Equations~(\ref{eq3}) and~(\ref{eq4}), $p=n_0+o(p)$, and thus
$p=O(n_0)$. Hence $\sum_{i\geq 1}\sum_{G\in S(i)}|G|^c=o(n_0)$. By
Equations~(\ref{eq2}) and~(\ref{eq3}), $p=n_0+o(n_0)$ and
$q=o(n_0)$, finishing the proof of Statement~1.

\paragraph{\rm Statement 2}
By Conditions~P\ref{s1} and P\ref{s2}, $|H|=\Omega(\lambda)$ for every
$H\in S(0)$. Since $\sum_{H\in S(0)}|H|=p=n_0+o(n_0)$,
$|S(0)|=O(n_0/\lambda)$. Together with Equation~(\ref{eq1}), we know
$|S(i)|=O(\frac{n_0}{r^i\lambda})$ for every $i=0,\ldots,\ell$. By the
definition of $S(i)$, $|G|\leq r^i\lambda$ for every
$i=0,\ldots,\ell$.  Therefore $G_0$ and $\algo{G_0}$ can be obtained
from each other in time
\begin{displaymath}
\frac{n_0}{\lambda}O(f(\lambda)+\sum_{1\leq i\leq\ell}r^{-i}g(r^i\lambda)).
\end{displaymath}
Clearly $f(\lambda)=2^{\lambda^{O(1)}}=2^{o(\log \log n_0)}=o(\log
n_0)$.  Since $\ell=O(\log n_0)$ and $g(n)=O(n)$, $\sum_{1\leq
i\leq\ell}r^{-i}g(r^i\lambda)=\sum_{1\leq i\leq\ell}\lambda=O(\lambda
\log n_0)$, and Statement~2 follows.
\end{proof}

Sections~\ref{section:triangulation} and~\ref{section:planar} use
Theorem~\ref{theorem:main} to encode various classes of graphs
$G$. Section~\ref{section:triangulation} considers plane
triangulations. Section~\ref{section:planar} considers planar graphs
and plane graphs.

\section{Plane triangulations}\label{section:triangulation}
A {\em plane triangulation} is a plane graph, each of whose faces has
size exactly three.  Note that a plane triangulation may contain self-loops and
multiple edges. Every $n$-node plane triangulation, simple or not, has
exactly $3n-6$ edges. In this section, let $\pi$ be an arbitrary
conjunction over the following groups of properties of a plane
triangulation $G$: F\ref{f2}, \Xomit{F\ref{f4}, F\ref{f5},} F\ref{f6},
and~F\ref{f7}, where $\ell_3$ and $\ell_4$ are not required to be
$O(1)$.
Our encoding scheme is based on the next fact.

\begin{fact}[See~\cite{Miller86}]\label{fact:separator}
Let $H$ be an $n$-node $m$-edge undirected plane
graph, each of whose faces has size at most $d$. We can compute a
node-simple cycle $C$ of $H$ in $O(n+m)$ time such that
\begin{itemize}
\item $C$ has at most $2\sqrt{dn}$ nodes; and
\item the numbers of $H$'s nodes inside and outside $C$ are at most
$2n/3$, respectively.
\end{itemize}
\end{fact}

Let $G$ be a given $n$-node $\pi$-graph.  Let $G'$ be obtained from
the undirected version of $G$ by deleting the self-loops. Clearly each
face of $G'$ has size at most four.  Let $C'$ be a cycle of $G'$
having size at most $4\sqrt{n}$ guaranteed by
Fact~\ref{fact:separator}.  Let $C$ consist of the edges of $G$
corresponding to the edges of $C'$ in $G'$.  Note that $C$ is not
necessarily a directed cycle if $G$ is directed. Since $G'$ does not
have self-loops, $2\leq|C|\leq 4\sqrt{n}$. If $\ell_4\geq 2$, then
$|C|$ can be two.  Let $\gin$ (respectively, $\gout$) be the
subgraph of $G$ formed by $C$ and the part of $G$ inside
(respectively, outside) $C$. Let $x$ be an arbitrary node on $C$.

$G_1$ is obtained by placing a cycle $C_1$ of three nodes outside
$\gin$ and then triangulating the face between $C_1$ and $\gin$ such
that a particular node $y_1$ of $C_1$ has degree strictly lower than
the other two. Clearly this is doable even if $|C|=2$. The
edge directions of $G_1-\gin$ can be
arbitrarily assigned according to $\pi$.

$G_2$ is obtained from $\gout$ by (1) placing a cycle $C_2$ of three
nodes outside $\gout$ and then triangulating the face between $C_2$
and $\gout$ such that a particular node $y_2$ of $C_2$ has degree
strictly lower than the other two; and (2) triangulating the face
inside $C$ by placing a new node $z$ inside of $C$ and then connecting
it to each node of $C$ by an edge. Note that (2) is doable even if
$|C|=2$. Similarly, the edge directions
of $G_2-\gout$ can be arbitrarily assigned according to $\pi$.

Let $u$ be a node of $G$. Let $v$ be a node on the boundary $B(G)$ of
the exterior face of $G$. Define $\dfs(u,G,v)$ as follows. Let $w$ be
the counterclockwise neighbor of $v$ on $B(G)$.  We perform a
depth-first search of $G$ starting from $v$ such that (1) the
neighbors of each node are visited in the counterclockwise order
around that node; and (2) $w$ is the second visited node.  A numbering
is assigned the first time a node is visited.  Let $\dfs(u,G,v)$ be
the binary number assigned to $u$ in the above depth-first search.
Let $X=\dfs(x,G_1,y_1)+\dfs(x,G_2,y_2)+\dfs(z,G_2,y_2)$.

\begin{lemma}\label{lemma:triangulation}
\begin{enumerate}
\item\label{i1} $G_1$ and $G_2$ are $\pi$-graphs.
\item\label{i2} There exists a constant $r>1$ with
$\max(|G_1|,|G_2|)\leq n/r$.
\item\label{i3} $|G_1|+|G_2|=n+O(\sqrt{n})$.
\item\label{i4} $|X|=O(\log n)$.
\item\label{i5} $G_1,G_2,X$ can be obtained from $G$ in $O(n)$ time.
\item\label{i6} $G$ can be obtained from $G_1,G_2,X$ in $O(n)$ time.
\end{enumerate}
\end{lemma}

\begin{proof}
Statements~\ref{i1}--\ref{i5} are straightforward by
Fact~\ref{fact:separator} and the definitions of $G_1$, $G_2$ and $X$.
Statement~\ref{i6} is proved as follows.
It takes $O(n)$ time to locate $y_1$ (respectively, $y_2$) in $G_1$
(respectively, $G_2$) by looking for the node with the lowest degree
on $B(G_1)$ (respectively, $B(G_2)$). By Fact~\ref{fact:aux}, it takes
$O(1)$ time to obtain $\dfs(y_1,G_1,x)$, $\dfs(y_2,G_2,x)$, and
$\dfs(y_2,G_2,z)$ from $X$. Therefore $x$ and $z$ can be located in
$G_1$ and $G_2$ in $O(n)$ time by depth-first traversal. Now $\gin$
can be obtained from $G_1$ by removing $B(G_1)$ and its incident
edges.  The cycle $C$ in $\gin$ is simply $B(\gin)$.  Also, $\gout$
can be obtained from $G_2$ by removing $B(G_2)$, $z$, and their
incident edges. The $C$ in $\gout$ is simply the boundary of the
face that encloses $z$ and its incident edges in $G_2$. Since we
know the positions of $x$ in $\gin$ and $\gout$, $G$ can be obtained
from $\gin$ and $\gout$ by fitting them together along $C$ by aligning
$x$. The overall time complexity is $O(n)$.
\end{proof}

\begin{theorem}\label{theorem:triangulation}
Let $G_0$ be an $n_0$-node $\pi$-graph. Then $G_0$ and $\algo{G_0}$
can be obtained from each other in $O(n_0\log n_0)$ time. Moreover,
$|\algo{G_0}|\leq\beta(n_0)+o(\beta(n_0))$ for any continuous
super-additive function $\beta(n)$ such that there are at most
$2^{\beta(n)+o(\beta(n))}$ distinct $n$-node $\pi$-graphs.
\end{theorem}
\begin{proof}
Since an $n$-node $\pi$-graph has $O(n)$ edges, there are at most
$2^{O(n\log n)}$ distinct $n$-node $\pi$-graphs.  Thus, there exists
an indexing scheme $\index{G}$ such that $\index{G}$ and $G$ can be
obtained from each other in $2^{|G|^{O(1)}}$ time. The theorem
follows from Theorem~\ref{theorem:main} and
Lemma~\ref{lemma:triangulation}.
\end{proof}

\begin{figure}
\begin{center}
\input{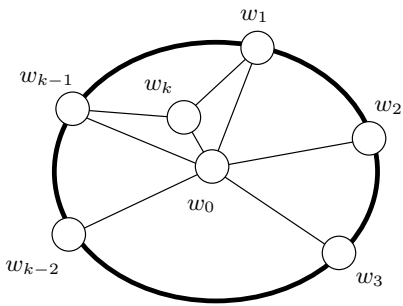}
\end{center}
\caption{A $k$-wheel graph $W_k$.\label{figure:wheel}}
\end{figure}

\section{Planar graphs and plane graphs}\label{section:planar}
In this section, let $\pi$ be an arbitrary conjunction over the
following groups of properties of $G$: F\ref{f1}, F\ref{f2},
F\ref{f3}, \Xomit{F\ref{f4}, F\ref{f5},} F\ref{f6}, and F\ref{f7}.
Clearly an $n$-node $\pi$-graph has $O(n)$ edges.

Let $G$ be an input $n$-node $\pi$-graph. For the cases of
$\pi$ being a planar graph rather than a plane graph, let $G$
be embedded first.  Note that this is only for the encoding process to
be able to apply Fact~\ref{fact:separator}. At the base level, we
still use the indexing scheme for $\pi$-graphs rather than the one
for embedded $\pi$-graphs. As shown below, the decoding
process does not require the $\pi$-graphs to be embedded.

Let $G'$ be obtained from the undirected version of $G$ by (1)
triangulating each of its faces that has size more than three such
that no additional multiple edges are introduced; and then (2)
deleting its self-loops.  Let $C'$ be a cycle of $G'$ guaranteed by
Fact~\ref{fact:separator}.  Let $C$ consists of the (a) edges of $G$
corresponds to the edges of $C'$ in $G'$, and (b) the edges of $C'$
that are added into $G'$ by the triangulation.  ($C$ is not
necessarily a directed cycle of a directed $G$.)  Let $G_C$ be the
union of $G$ and $C$.  Let $\gin$ (respectively, $\gout$) be the
subgraph of $G_C$ formed by $C$ and the part of $G_C$ inside
(respectively, outside) $C$. Let $C=x_1x_2\cdots x_\ell x_{\ell+1}$,
where $x_{\ell+1}=x_1$.  By Fact~\ref{fact:separator},
$\ell=O(\sqrt{n})$.
\begin{lemma}
\label{lemma:degree}
Let $H$ be an $O(n)$-node $O(n)$-edge graph. There exists an
integer $k$ with $n^{0.6}\leq k\leq n^{0.7}$ such that $H$ does not
contain any node of degree $k$ or $k-1$.
\end{lemma}
\begin{proof}
Assume for a contradiction that such a $k$ does not exist. It follows
that the sum of degrees of all nodes in $H$ is at least
$(n^{0.6}+n^{0.7})(n^{0.7}-n^{0.6})/4=\Omega(n^{1.4})$. This
contradicts the fact that $H$ has $O(n)$ edges.
\end{proof}

Let $W_k$, with $k\geq3$, be a {\em $k$-wheel graph} defined as
follows. As shown in Figure~\ref{figure:wheel}, $W_k$ consists of
$k+1$ nodes $w_0,w_1,w_2,\ldots,w_{k-1},w_k$, where
$w_1,w_2,\ldots,w_k,w_1$ form a cycle.  $w_0$ is a degree-$k$ node
incident to each node on the cycle. Finally, $w_1$ is incident to
$w_{k-1}$.  Clearly $W_k$ is triconnected. Also, $w_1$ and $w_k$ are the
only degree-four neighbors of $w_0$ in $W_k$.  Let $k_1$
(respectively, $k_2$) be an integer $k$ guaranteed by
Lemma~\ref{lemma:degree} for $\gin$ (respectively, $\gout$). Now we
define $G_1$, $G_2$ and $X$ as follows.

\begin{figure}
\begin{center}
\input{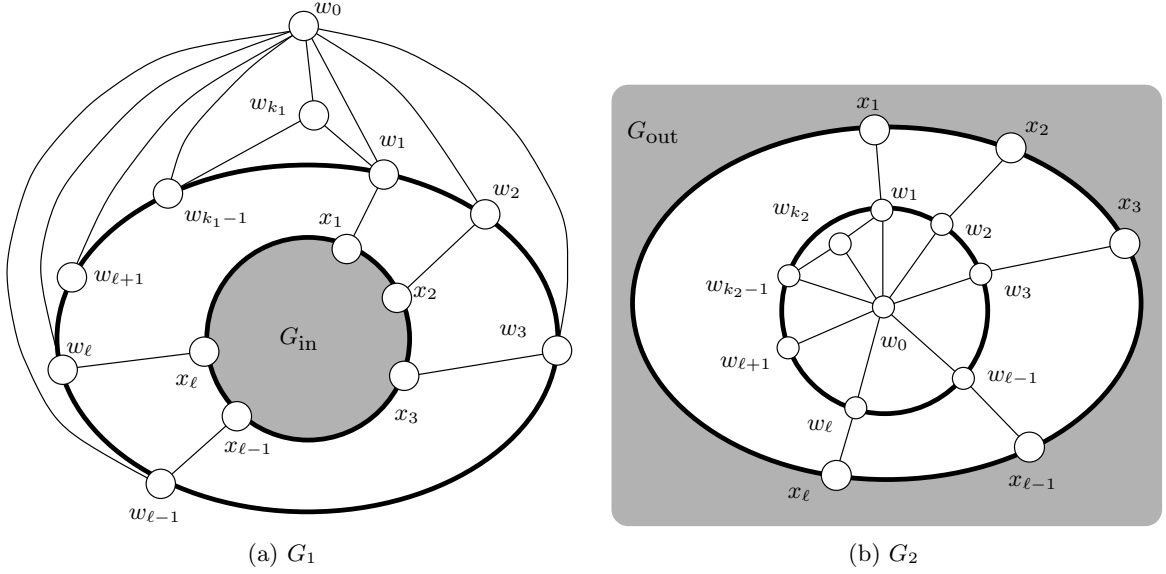}
\end{center}
\caption{$G_1$ and $G_2$. The gray area of $G_1$ is $\gin$. The gray
area of $G_2$ is $\gout$.\label{figure:planar}}
\end{figure}

$G_1$ is obtained from $\gin$ and a $k_1$-wheel graph $W_{k_1}$ by
adding an edge $(w_i,x_i)$ for every $i=1,\ldots,\ell$. Clearly for
the case of $\pi$ being a plane graph, $G_1$ can be embedded
such that $W_{k_1}$ is outside $\gin$, as shown in
Figure~\ref{figure:planar}(a). Thus, the original embedding of $\gin$
can be obtained from $G_1$ by removing all nodes of $W_{k_1}$.  The
edge directions of $G_1-\gin$ can be
arbitrarily assigned according to $\pi$.

$G_2$ is obtained from $\gout$ and a $k_2$-wheel graph $W_{k_2}$ by
adding an edge $(w_i,x_i)$ for every $i=1,\ldots,\ell$. Clearly for
the case of $\pi$ being a plane graph, $G_2$ can be embedded
such that $W_{k_2}$ is inside $C$, as shown in
Figure~\ref{figure:planar}(b). Thus, the original embedding of
$\gout$ can be obtained from $G_2$ by removing all nodes of $W_{k_2}$.
The edge directions of $G_2-\gout$ can
be arbitrarily assigned according to $\pi$.

Let $X$ be an $O(\sqrt{n})$-bit string which encodes $k_1$, $k_2$,
and whether each edge $(x_i,x_{i+1})$ is an original edge in $G$, for
$i=1,\ldots,\ell$. 

\begin{lemma}
\label{lemma:planar}
\begin{enumerate}
\item\label{t1} $G_1$ and $G_2$ are $\pi$-graphs.
\item\label{t2} There exists a constant $r>1$ with $\max(|G_1|,|G_2|)\leq
n/r$.
\item $|G_1|+|G_2|=n+O(n^{0.7})$.
\item $|X|=O(\sqrt{n})$.
\item\label{t3} $G_1,G_2,X$ can be obtained from $G$ in $O(n)$ time.
\item\label{t4} $G$ can be obtained from $G_1,G_2,X$ in $O(n)$ time.
\end{enumerate}
\end{lemma}
\begin{proof}
Since $W_{k_1}$ and $W_{k_2}$ are both triconnected, and each node of
$C$ has degree at least three in $G_1$ and $G_2$, Statement~\ref{t1}
holds for each case of the connectivity of the input $\pi$-graph $G$.
Statements~\ref{t2}--\ref{t3} are straightforward by
Fact~\ref{fact:separator} and the definitions of $G_1$, $G_2$ and $X$.
Statement~\ref{t4} is proved as follows. First of all, we obtain $k_1$
from $X$. Since $\gin$ does not contain any node of degree $k_1$ or
$k_1-1$, $w_0$ is the only degree-$k_1$ node in $G_1$. Therefore it
takes $O(n)$ time to identify $w_0$ in $G_1$. $w_{k_1}$ is the only
degree-3 neighbor of $w_0$. Since $k_1>\ell$, $w_1$ is the only
degree-5 neighbor of $w_0$. $w_2$ is the common neighbor of
$w_0$ and $w_1$ that is not adjacent to $w_{k_1}$.  From now on,
$w_i$, for each $i=3,4,\ldots,\ell$, is the common neighbor of $w_0$
and $w_{i-1}$ other than $w_{i-2}$. Clearly, $w_1,w_2,\ldots,w_\ell$
and thus $x_1,x_2,\ldots,x_\ell$ can be identified in $O(n)$
time. $\gin$ can now be obtained from $G_1$ by removing
$W_{k_1}$. Similarly, $\gout$ can be obtained from $G_2$ and $X$ by
deleting $W_{k_1}$ after identifying
$x_1,x_2,\ldots,x_{\ell}$. Finally, $G_C$ can be recovered by fitting
$\gin$ and $\gout$ together by aligning $x_1,x_2,\ldots,x_\ell$. Based
on $X$, $G$ can then be obtained from $G_C$ by removing the edges of
$C$ that are not originally in $G$.
\end{proof}

\paragraph{Remark} In the proof for Statement~\ref{t4} 
of Lemma~\ref{lemma:planar}, identifying the degree-$k_1$ node (and
the $k_1$-wheel graph $W_{k_1}$) does not require the embedding for
$G_1$. Therefore the decoding process does not require the
$\pi$-graphs to be embedded. This is different from the proof of
Lemma~\ref{lemma:triangulation}.

\begin{theorem}\label{theorem:planar}
Let $G_0$ be an $n_0$-node $\pi$-graph. Then $G_0$ and $\algo{G_0}$
can be obtained from each other in $O(n_0\log n_0)$ time. Moreover,
$|\algo{G_0}|\leq\beta(n_0)+o(\beta(n_0))$ for any continuous
super-additive function $\beta(n)$ such that there are at most
$2^{\beta(n)+o(\beta(n))}$ distinct $n$-node $\pi$-graphs.
\end{theorem}
\begin{proof}
Since there are at most $2^{O(n\log n)}$ distinct $n$-node
$\pi$-graphs, there exists an indexing scheme $\index{G}$ such that
$\index{G}$ and $G$ can be obtained from each other in
$2^{|G|^{O(1)}}$ time. 
The theorem follows from Theorem~\ref{theorem:main} and
Lemma~\ref{lemma:planar}.
\end{proof}

\section{Concluding remarks}\label{sec_open}
For brevity, we left out F\ref{f4} and F\ref{f5} in
Sections~\ref{section:triangulation} and~\ref{section:planar}. One can
verify that Theorems~\ref{theorem:triangulation} and~\ref{theorem:planar}
hold even if $\pi$ is a conjunction over F\ref{f4} and F\ref{f5}.

The coding schemes given in this paper require $O(n\log n)$ time
for encoding and decoding. An immediate open question is whether one
can encode some graphs other than rooted trees in $O(n)$ time using
information-theoretically minimum number of bits. It would be of
significance to determine whether the tight bound of the
number of distinct $\pi$-graphs for each $\pi$ is indeed continuous
super-additive.

\bibliographystyle{siam}
\bibliography{encode}

\begin{thebibliography}{10}

\bibitem{BCW90}
{\sc T.~C. Bell, J.~G. Cleary, and I.~H. Witten}, {\em Text Compression},
  Prentice-Hall, Englewood Cliffs, NJ, 1990.

\bibitem{ChuangGHKL98}
{\sc R.~C.-N. Chuang, A.~Garg, X.~He, M.-Y. Kao, and H.-I. Lu}, {\em Compact
  encodings of planar graphs via canonical orderings and multiple parentheses},
  in Automata, Languages and Programming, 25th Colloquium, K.~G. Larsen,
  S.~Skyum, and G.~Winskel, eds., vol.~1443 of Lecture Notes in Computer
  Science, Aalborg, Denmark, 13--17~July 1998, Springer-Verlag, pp.~118--129.

\bibitem{Elias75}
{\sc P.~Elias}, {\em Universal codeword sets and representations of the
  integers}, {IEEE} Transactions on Information Theory, IT-21 (1975),
  pp.~194--203.

\bibitem{GW83}
{\sc H.~Galperin and A.~Wigderson}, {\em Succinct representations of graphs},
  Information and Control, 56 (1983), pp.~183--198.

\bibitem{HeKL-sidm}
{\sc X.~He, M.-Y. Kao, and H.-I. Lu}, {\em Linear-time succinct encodings of
  planar graphs via canonical orderings}, {SIAM} Journal on Discrete
  Mathematics,  (1999).
\newblock To appear.

\bibitem{IR82}
{\sc A.~Itai and M.~Rodeh}, {\em Representation of graphs}, Acta Informatica,
  17 (1982), pp.~215--219.

\bibitem{Jacobson89}
{\sc G.~Jacobson}, {\em Space-efficient static trees and graphs}, in 30th
  Annual Symposium on Foundations of Computer Science, Research Triangle Park,
  North Carolina, 30 Oct.--1 Nov. 1989, IEEE, pp.~549--554.

\bibitem{KNR92}
{\sc S.~Kannan, N.~Naor, and S.~Rudich}, {\em Implicit representation of
  graphs}, {SIAM} Journal on Discrete Mathematics, 5 (1992), pp.~596--603.

\bibitem{kaot93.joa}
{\sc M.~Y. Kao, N.~Occhiogrosso, and S.~H. Teng}, {\em Simple and efficient
  compression schemes for dense and complement graphs}, Journal of
  Combinatorial Optimization, 2 (1999), pp.~351--359.

\bibitem{KW:encodings}
{\sc K.~Keeler and J.~Westbrook}, {\em Short encodings of planar graphs and
  maps}, Discrete Applied Mathematics, 58 (1995), pp.~239--252.

\bibitem{LiptonT79}
{\sc R.~J. Lipton and R.~E. Tarjan}, {\em A separator theorem for planar
  graphs}, {SIAM} Journal on Applied Mathematics, 36 (1979), pp.~177--189.

\bibitem{Miller86}
{\sc G.~L. Miller}, {\em Finding small simple cycle separators for 2-connected
  planar graphs}, Journal of Computer and System Sciences, 32 (1986),
  pp.~265--279.

\bibitem{MR97}
{\sc J.~I. Munro and V.~Raman}, {\em Succinct representation of balanced
  parentheses, static trees and planar graphs}, in 38th Annual Symposium on
  Foundations of Computer Science, Miami Beach, Florida, 20--22 Oct. 1997,
  IEEE, pp.~118--126.

\bibitem{naor90}
{\sc M.~Naor}, {\em Succinct representation of general unlabeled graphs},
  Discrete Applied Mathematics, 28 (1990), pp.~303--307.

\bibitem{PH86.encode}
{\sc C.~H. Papadimitriou and M.~Yannakakis}, {\em A note on succinct
  representations of graphs}, Information and Control, 71 (1986), pp.~181--185.

\bibitem{turan84}
{\sc G.~Tur\'{a}n}, {\em On the succinct representation of graphs}, Discrete
  Applied Mathematics, 8 (1984), pp.~289--294.

\bibitem{Tutte62}
{\sc W.~T. Tutte}, {\em A census of planar triangulations}, Canadian Journal of
  Mathematics, 14 (1962), pp.~21--38.

\bibitem{Tutte63b}
\leavevmode\vrule height 2pt depth -1.6pt width 23pt, {\em A census of planar
  maps}, Canadian Journal of Mathematics, 15 (1963), pp.~249--271.

\end{thebibliography}

\end{document}